\documentclass[twocolumn,showpacs,preprintnumbers,amsmath,amssymb]{revtex4}

\usepackage{graphicx}
\usepackage{dcolumn}
\usepackage{bm}

\begin{document}

\preprint{APS/123-QED}

\title{Standard Nearest Neighbor Discretizations of Klein-Gordon Models \\ Cannot Preserve Both Energy and Linear Momentum}

\author{S. V. Dmitriev$^1$, P. G. Kevrekidis$^2$ and N. Yoshikawa$^3$}
\affiliation{
$^1$ National Institute of Materials Science, 1-2-1 Sengen, Tsukuba, Ibaraki 305-0047, Japan \\
$^2$ Department of Mathematics and Statistics, University of Massachusetts,
Amherst, MA 01003-4515, USA  \\
$^3$ Institute of Industrial Science, University of Tokyo, Komaba,
Meguro-ku, Tokyo 153-8505, Japan }
\date{\today}

\begin{abstract}
We consider  nonlinear Klein-Gordon wave equations and illustrate
that standard discretizations thereof (involving nearest
neighbors) may preserve either standardly defined linear momentum
or total energy but not both. This has a variety of intriguing
implications for the ``non-potential'' discretizations that
preserve only the linear momentum, such as the self-accelerating
or self-decelerating motion of coherent structures such as
discrete kinks in these nonlinear lattices.
\end{abstract}


\maketitle

{\it Introduction}.
In the last two decades, the interplay of nonlinearity and spatial
discreteness has been increasingly recognized as vital for the understanding
of a variety of physical systems \cite{reviews}. Such contexts
range from calcium waves in living
cells \cite{g1} to the propagation of action potentials through
the cardiac tissue  \cite{g2} and from chains of
chemical reactions \cite{g3} to applications in superconductivity
and Josephson junctions \cite{g4}, nonlinear optics and fiber/waveguide
arrays \cite{g5}, complex electronic materials \cite{g6}, Bose-Einstein
condensates \cite{ts} or the
local denaturation of the DNA double strand \cite{g7}.

On the other hand, spatially discrete systems (of coupled
nonlinear ordinary differential equations) are also relevant as
discretizations and computational implementations of the
corresponding continuum field theories that are applicable to a
variety of contexts such as statistical mechanics \cite{parisi},
solid state physics \cite{eilbeck}, fluid mechanics \cite{infeld}
and particle physics \cite{phi4} (see also references therein).
Nonlinear Klein-Gordon type equations are a prototypical example
among such wave models and their variants span a variety of
applications including Josephson junctions in superconductivity,
cosmic domain walls in cosmology, elementary particles in
particle physics and denaturation bubbles in the
DNA, among others.

In this communication, we examine some of the key properties that
ensue when discretizing nonlinear Klein-Gordon (KG) equations,
using nearest neighbor approximations (which are the most standard
ones implemented in the literature; see e.g., \cite{reviews}). In
particular, we focus on the physically relevant invariances of the
continuum equation (more specifically, the conservation of the
linear momentum and of the total energy of the system) and
illustrate the surprising result that if we demand that the energy
be conserved, then the momentum cannot be conserved, while if we
demand that the momentum be conserved then the energy cannot be
conserved (resulting in a so-called non-potential model
\cite{bolotin}). Our presentation will be structured as follows.
First, we will provide the general mathematical setting of KG
equations and study their discretizations that conserve linear
momentum and energy, comparing and constructing the properties of
the two. Then, we are going to give an application of our
considerations to the physically relevant $\phi^4$ model, i.e.,
the ubiquitous double well potential. Finally, we will summarize
our conclusions and discuss future directions.

{\it Setup and Analytical Results}. We consider the Lagrangian of
the Klein-Gordon field,
\begin{equation}
L =\int_{-\infty}^{\infty} \left[
\frac{1}{2}\phi_t^2-\frac{1}{2}\phi_x^2-V(\phi)\right]dx\,,
\label{KleinGordonHam}
\end{equation}
and the corresponding equation of motion,
\begin{equation}
\phi _{tt}  = \phi _{xx}  - V'(\phi)\,. \label{KleinGordon}
\end{equation}
Assuming that the background potential $V(\phi)$ can be expanded
in Taylor series we write
\begin{equation}
V'\left( \phi  \right) = \sum_{s=0}^{\infty} \sigma _s \phi^s \,.
\label{PolynomPotential}
\end{equation}
For brevity, when possible, we will use the notations
\begin{equation}
\phi _{n - 1}  \equiv l,\,\,\,\,\,\phi _n  \equiv m,\,\,\,\,\,\phi
_{n + 1}  \equiv r\,. \label{Notation}
\end{equation}

We start with a general proof of our main statement, namely that
discretizations that preserve linear momentum and energy are
mutually exclusive for nearest neighbor discretizations. As was
shown in \cite{PhysicaD}, the standard discretization of Eq.
(\ref{KleinGordon}) that preserves the discrete analog of the
linear momentum, defined in a standard way,
\begin{eqnarray}
M= \sum_{n=-\infty}^{\infty} \dot{\phi}_n
\left(\phi_{n+1}-\phi_{n-1} \right), \label{mom1}
\end{eqnarray}
is one of the form:
\begin{eqnarray}
\ddot{m}=C \left(l+r-2 m \right) - \frac{F(r,m)-F(m,l)}{r-l},
\label{mom2}
\end{eqnarray}
where $C=1/h^2$, where $h$ is the lattice spacing, and the
derivative of $F$ is equal to $V$ in the continuum limit
$(C\rightarrow \infty)$. Then,
\begin{eqnarray}
\frac{dM}{dt}=\sum_n \ddot{\phi}_n
(\phi_{n+1}-\phi_{n-1})\nonumber \\
= \sum_n [H(\phi_{n+1},\phi_{n}) - H(\phi_{n},\phi_{n-1})]=0,
\label{mom3}
\end{eqnarray}
where $H(r,m)=C (r^2+m^2-2 m r) - F(r,m)$, and the terms
$\dot{\phi}_n(\dot{\phi}_{n+1}-\dot{\phi}_{n-1})$ cancel out as
the telescopic sum.

However, if the model is potential, for nearest neighbor
discretizations the nonlinear term will be of the form
$\tilde{V}(r,m)$ such that the Lagrangian can be written as:
\begin{eqnarray}
L=\sum_n \left[\frac{1}{2}  \dot{\phi}_n^2 - \frac{C}{2}
(\phi_{n+1}-\phi_n)^2 - \tilde{V}(\phi_{n+1},\phi_{n}) \right],
\label{mom4}
\end{eqnarray}
where the first term gives the kinetic energy, $ {\cal K}$, and the two
other terms give negative potential energy, $-{\cal P}$, so that the
total energy is $E={\cal K}+ {\cal P}$. However, then a model that
would enforce
both energy and momentum conservation would have to satisfy:
\begin{equation}
\frac{F(r,m)-F(m,l)}{r-l} = \frac{\partial}{\partial m}\left[
\tilde{V}(r,m) + \tilde{V}(m,l)\right]. \label{mom5}
\end{equation}
After multiplying with $r-l$, this, in turn, implies that the
cross terms involving all 3 of $r,m$ and $l$ should be presentable
in the form
\begin{eqnarray}
r \frac{\partial
\tilde{V}(m,l)}{\partial m} - l \frac{\partial \tilde{V}(r,m)}{\partial m}=
P(r,m)-P(m,l). \label{mom6}
\end{eqnarray}
This is satisfied only if $\tilde{V}(x,y)$ is a (symmetric)
quadratic function in its arguments. However, this is incompatible
with the nonlinear nature of the model. Hence, it is not possible
to satisfy both conservation laws at once.

Let us now derive the general discrete Klein-Gordon model of the form
of Eq. (\ref{mom2}) conserving momentum. For the polynomial
background forces Eq. (\ref{PolynomPotential}), the nonlinear term
of Eq. (\ref{mom2}) can be presented as the sum of $s$-order terms
\begin{equation}
B(l,m,r)=\frac{F(r,m)-F(m,l)}{r-l} = \sum_{s=0}^{\infty} B_s
(l,m,r)\,, \label{BasSum}
\end{equation}
with
\begin{eqnarray}
B_s  = \sum\limits_{i = 0}^s {\sum\limits_{j = i}^s {b_{ij,s} r^i
m^{j-i} l^{s - j} } }, \label{Bterms}
\end{eqnarray}
where
\begin{equation}
\sum\limits_{i = 0}^s {\sum\limits_{j = i}^s {b_{ij,s} }}=\sigma
_s\,\,. \label{CoeffConditions}
\end{equation}
In the continuum limit one has $l\rightarrow m$ and $r\rightarrow
m$ and thus, Eq. (\ref{BasSum}) together with Eq.
(\ref{CoeffConditions}) ensure the desired limit, $V'(\phi)$.
Furthermore, Eq. (\ref{Bterms}) takes into account all possible
combinations of powers of $l,m,$ and $r$. Coefficients $b_{ij,s}$
make a triangular matrix of size $(s+1)\times (s+1)$. Let us find
the coefficients $b_{ij,s}$ to satisfy Eq. (\ref{BasSum}). We
write
\begin{eqnarray}
(r - l)B_s  = \sum\limits_{i = 0}^s {\sum\limits_{j = i}^s
{b_{ij,s} r^{i + 1} m^{j-i} l^{s - j} } } \nonumber \\
- \sum\limits_{i = 0}^s {\sum\limits_{j = i}^s {b_{ij,s} r^i
m^{j-i} l^{s - j + 1} } }. \label{BsD1a}
\end{eqnarray}
Terms containing both $l$ and $r$ should be canceled out because
they do not fit the representation of Eq. (\ref{BasSum}). This can be
achieved by setting $b_{ij,s} = b_{(i + 1)(j + 1),s} $ , i.e.,
coefficients in each diagonal of the triangular matrix must be
equal. The simplified expression reads:
\begin{eqnarray}
(r - l)B_s  = \sum\limits_{i = 0}^s {b_{is,s} r^{i + 1} m^{s - i}
} - \sum\limits_{i = 0}^s {b_{0i,s} m^i l^{s - i + 1} }.
\label{BsD1b}
\end{eqnarray}
To symmetrize the result, we add and subtract $b_{00,s} m^{s+1} $
\begin{eqnarray}
(r - l)B_s  = b_{00,s} (r^{s+1}  + m^{s+1} ) - b_{00,s} (m^{s+1}  + l^{s+1} ) \nonumber \\
+\sum\limits_{i = 1}^{s} {b_{0(s-i+1),s} r^i m^{s - i + 1}}
-\sum\limits_{i = 1}^s {b_{0i,s} m^i l^{s - i + 1} },
\label{BsD1c}
\end{eqnarray}
where we shifted the summation index by 1 in the first sum and
also used the equality of the diagonal coefficients. The desired
representation is obtained for arbitrary $b_{00,s}$ and arbitrary
$b_{0i,s} = b_{0(s-i+1),s}$ for $i>0$. Summing up, (i) the
coefficients $b_{ij,s}$ within each diagonal are equal, (ii) the
coefficients on the main diagonal can be chosen arbitrarily, and
(iii) the terms on $i$th super-diagonal $(i>0)$ must have the same
coefficients as the terms on $(s - i + 1)$th diagonal (and these
can also be chosen arbitrarily).

For $B_s$ the number of super-diagonals is $s$ so that the number
of free coefficients is $1 + \left\lceil {s/2} \right\rceil$,
where $\left\lceil x \right\rceil$ is lowest integer greater than
or equal to $x$. We must also take into account the relation
between coefficients Eq. (\ref{CoeffConditions}) and the number of
free coefficients becomes $\left\lceil {s/2} \right\rceil$.

For example, the coefficients of $B_3$ are
\begin{eqnarray}
b_{ij,3} = \left[ {\begin{array}{*{20}c}
{b_{00,3} } & {b_{01,3} } & {b_{02,3} } & {b_{01,3} }  \\
{} & {b_{00,3} } & {b_{01,3} } & {b_{02,3} }  \\
{} & {} & {b_{00,3} } & {b_{01,3} }  \\
{} & {} & {} & {b_{00,3} }  \\
\end{array} } \right], \nonumber \\
4b_{00,3} + 4b_{01,3} +2b_{02,3} = \sigma_3. \label{B1B3D1}
\end{eqnarray}

Since the model Eq. (\ref{mom2}) is translationally invariant, the
static kink is free of the Peierls-Nabarro potential (PNp)
\cite{PhysicaD}, i.e., the periodic potential that nonlinear waves
have to overcome to move by one lattice site \cite{malomedkivshar}
(see also references therein). This  is an important qualitative
difference with respect to the conventional discretization when,
in Eq. (\ref{KleinGordonHam}), $V(\phi)$ is substituted with
$V(\phi_n)$ and thus, in the equation of motion Eq.
(\ref{KleinGordon}), $V'(\phi)$ becomes $V'(\phi_n)$.

Another class of Klein-Gordon models which support energy conservation
and sustain static kinks but which are free of PNp
has been derived by Speight and collaborators \cite{SpeightKleinGordon}.
In such models the background potential term of Eq.
(\ref{KleinGordonHam}), $P=\int V(\phi)dx$, should be discretized
as
\begin{equation}
P= \sum_n
\left(\frac{G(\phi_{n+1})-G(\phi_{n})}{\phi_{n+1}-\phi_{n}}\right)^2,\,\,
{\rm with} \,\,\left[G'(\phi)\right]^2 = V(\phi)\,.
\label{Speight}
\end{equation}

{\it Numerical Results/Model Comparison}.
We now examine various models proposed as discretizations
of the continuum field theory in the context of perhaps one of
the most famous such examples, namely the double-well $\phi^4$ model
\cite{parisi,eilbeck,phi4} (see also the review \cite{belova}).

The discrete Klein-Gordon model conserving momentum is given by
Eq. (\ref{mom2}) with the nonlinear term Eq. (\ref{BasSum}) where
the coefficients $b_{ij,s}$ are as described in the previous
section. The continuum $\phi ^4 $ model has the background
potential $V(\phi)=(1 - \phi^2)^2/4$, hence
$V'(\phi)=-\phi+\phi^3$ so that in Eq. (\ref{PolynomPotential})
all $\sigma_s=0$ except for $\sigma_1=-\sigma_3=-1$.  The
momentum-preserving discretization then reads:
\begin{eqnarray}
\ddot m  = \left(C+\alpha\right)(l + r - 2m) +  m \, \nonumber \\
-\beta(l^2+lr+r^2)+\beta m(l+r+m) \, \nonumber \\
- \gamma \left( l^3 + r^3 + l^2r + lr^2 \right) -\delta m\left(
l^2 +
m^2 + r^2 + lr \right)\nonumber \\
- \frac{1}{2}(1-4\gamma-4\delta)m^2 \left( l+r \right) ,
\label{D1PHI4}
\end{eqnarray}
where $\alpha=-b_{00,1}$, $\beta=b_{00,2}$, $\gamma=b_{00,3}$,
$\delta=b_{01,3}$ are free parameters and we did not include the
terms with $s>3$.

The model of Eq. (\ref{D1PHI4}) will be compared to the model
obtained from Eq. (\ref{Speight}) in $\phi^4$ case
\cite{SpeightKleinGordon}, namely
\begin{eqnarray}
\ddot{m}=\left(C+\frac{1}{6}\right)(l + r - 2m)+m \nonumber \\
-\frac{1}{18}\left[2m^3+(m+l)^3+(m+r)^3\right]
,\label{PHI4Speight}
\end{eqnarray}
and also to the ``standard'' $\phi^4$ discretization, i.e.,
\begin{eqnarray}
\ddot{m}=C(l + r - 2m)+m -m^3. \label{PHI4Classic}
\end{eqnarray}

If in Eq. (\ref{D1PHI4}), $\alpha=\beta=\gamma=\delta=0$, then the
models of Eq. (\ref{D1PHI4}) and Eq. (\ref{PHI4Speight}) have the
same linear vibration spectrum (i.e., dispersion relation
$\omega=\omega(\kappa)$) for the vacuum solution $\phi_n=\pm 1$,
namely $\omega^2=2+(4C-2)\sin^2(\kappa/2)$. This can be compared
to the spectrum of the vacuum of Eq. (\ref{PHI4Classic}),
$\omega^2=2+4C\sin^2(\kappa/2)$.

We analyze the kink internal modes (i.e., internal degrees of
freedom \cite{Peli}) for these three models. First, we determine
the kink-like heteroclinic solution by means of relaxational
dynamics. Then, the linearized equations are used in a lattice of
$N=200$ sites to obtain $N$ eigenfrequencies and the corresponding
eigenmodes. We are particularly interested in the eigenfrequencies
which lie outside the linear vibration band of vacuum solution and
thus are associated with the kink internal modes. It is worthwhile
to notice that the eigenproblem for models conserving energy, Eq.
(\ref{PHI4Speight}) and Eq. (\ref{PHI4Classic}), has a symmetric
Hessian matrix while the non-self-adjoint problem for the momentum
conserving model Eq. (\ref{D1PHI4}) results in a {\it
non-symmetric} matrix.

\begin{figure}
\includegraphics{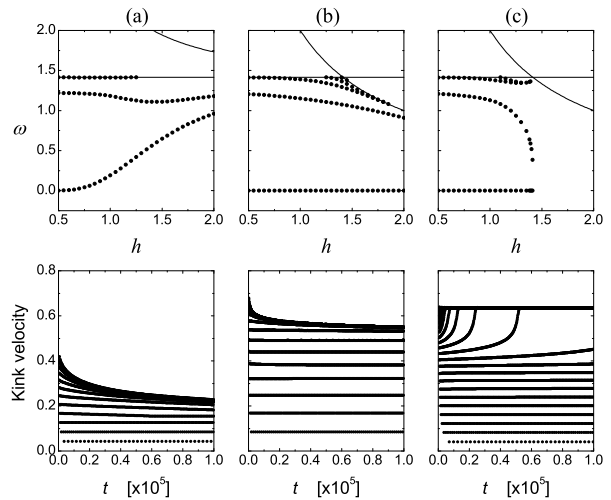}
\caption{Upper panels: boundaries of the linear spectrum of the
vacuum (solid lines) and kink internal mode frequencies (dots) as
functions of the lattice spacing $h=1/\sqrt{C}$. Lower panels:
time evolution of kink velocity for different initial velocities
and $h=0.7$. The results are shown for (a) classical $\phi^4$
model, Eq. (\ref{PHI4Classic}), (b) PNp-free model conserving
energy, Eq. (\ref{PHI4Speight}), and (c) PNp-free model conserving
momentum, Eq. (\ref{D1PHI4}), with
$\alpha=\beta=\gamma=\delta=0$.} \label{Figure1}
\end{figure}

The top panels of Fig. \ref{Figure1} present the boundaries of the
linear vibration spectrum of the vacuum (solid lines) and the kink
internal modes (dots) as the functions of lattice spacing $h$ for
(a) the classical $\phi^4$ model of Eq. (\ref{PHI4Classic}), (b)
the PNp-free model of Eq. (\ref{PHI4Speight}) conserving energy,
and (c) the PNp-free model of Eq. (\ref{D1PHI4}) conserving
momentum. In PNp-free models kinks possess a zero frequency,
Goldstone translational mode similarly to the continuum $\phi^4$
kink. Hence, the static kink can be centered anywhere on the
lattice. The results presented in Fig. \ref{Figure1} are for the
kink situated exactly between two lattice sites. This position is
the stable position for the classical $\phi^4$ discrete kink
\cite{Peli}. Since all $3$ discrete models share the same
continuum ($\phi^4$) limit, their spectra are very close for small
$h(<0.5)$. We found that the model Eq. (\ref{D1PHI4}) may have
kink internal modes lying {\em above} the spectrum of vacuum,
e.g., for $\alpha=1/2$, $\beta=0$, $\gamma=1/4$, and $\delta=0$.
Such modes are short-wavelength ones, with large amplitudes
(energies) and they do not radiate because of the absence of
coupling to the linear phonon spectrum.

\begin{figure}
\includegraphics{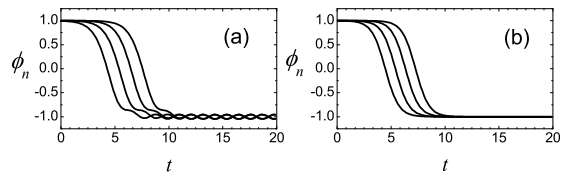}
\caption{Trajectories of particles (a) in the model of Eq.
(\ref{D1PHI4}) with $h=0.7$ when the kink moves with a steady
velocity $v^*$ (see Fig. \ref{Figure1}(c), bottom panel) and (b)
for the continuum $\phi^4$ kink.
} \label{Figure2}
\end{figure}

Perhaps more interesting are the implications of such
discretizations on the mobility of these localized coherent
structures. In the PNp-free models, Eq. (\ref{D1PHI4}) and Eq.
(\ref{PHI4Speight}), the kink was launched using a perturbation
along the Goldstone mode to provide the initial kick. In the
classical model Eq. (\ref{PHI4Classic}) for this purpose we
employed the imaginary frequency (real eigenvalue) unstable
eigenmode for a kink initialized at the unstable position (a
``site-centered'' kink). In all cases the amplitude of the mode is
related to the initial velocity of the kink. In the bottom panels
of Fig. \ref{Figure1} we present the time evolution of the kink
velocity for different initial velocities and $h=0.7$ for the
three discretizations. The results suggest that the mobility of
the kink in the classical $\phi^4$ model presented in (a) is much
smaller than in the PNp-free models, (b) and (c). Furthermore, a
very interesting effect of kink {\it self-acceleration} can be
observed in panel (c). Here there exists a selected kink velocity
$v^*\approx 0.637$ and kinks launched with $v>v^*$, in a very
short time adjust their velocities to $v^*$. More surprisingly,
the velocity adjustment is observed even for kinks launched with
$v<v^*$. In the steady-state regime, when the kink moves with
$v=v^*$, it excites (in its tail) the short-wave oscillatory mode
even though in front of the kink the vacuum is not perturbed.

These results generate the question of where the energy for the
self-acceleration and vacuum excitation comes from. In Fig. 2(a)
we show the trajectories of four neighboring particles when a kink
moving with $v=v^*$ (see Fig. \ref{Figure1}(c), bottom panel)
moves through. For comparison, in (b) the trajectories for the
classical $\phi^4$ kink, $\phi_n(t)=\tanh[\rho(nh-vt)]$, where
$\rho=[2-2v^2]^{-1/2}$, are shown. In both cases the trajectories
are identical and shifted with respect to each other by $t=h/v$,
but in (b) they are the odd functions with respect the point
$\phi_n=0$ while in (a) they are not. The work done by the
background forces, Eq. (\ref{BasSum}), to move the $n$th particle
from one energy well to another is
$W_n=-\int_{-\infty}^{\infty}\dot{m}B(l,m,r)dt$. For the $\phi^4$
model Eq. (\ref{D1PHI4}) with $\beta=\gamma=\delta=0$, the
nonlinear part of $B(l,m,r)$ reduces to $B(l,m,r)=(1/2)m^2(l+r)$.
It is straightforward to demonstrate that $W_n=0$ for the
classical $\phi^4$ kink. However, if a term breaking odd symmetry,
e.g., $\varepsilon\cosh^{-1}[\theta(nh-vt)]$, is added to the
kink, the work becomes nonzero,
$W_n=\frac{\pi}{2}\varepsilon(\varepsilon^2+1)[\cosh(\rho
h)-1]^3/\sinh^4(\rho h)$, where we set for simplicity
$\theta=\rho$. Numerically we found that $W_n$ can be positive or
negative depending on $\rho$, $\theta$ and the kink velocity, $v$.
This simple analysis qualitatively explains the  kink
self-acceleration or deceleration and the vacuum excitation. The
energy for this comes from the breaking of the odd symmetry of
particle trajectories, which is possible in the case of
path-dependent background forces. It is, thus, very interesting to
highlight the distinctions between the ``regular'' discrete
models, the PNp-free, energy conserving discrete models, and the
PNp-free, momentum conserving discrete models. The first ones lead
to rapid dissipation of the wave's kinetic energy due to the PN
barrier. The second render the dissipation far slower in time.
Finally the third may even sustain self-accelerating waves and
locking to a particular speed due to the non-potential nature of
the relevant model.

{\it Conclusions and Future Challenges}. The statement that
Klein-Gordon discrete model cannot conserve energy and momentum
simultaneously was proved for the case of standard nearest-neighbor
discretizations. This raises the issue of how to gauge the ``adequacy''
of a discretization scheme with respect to the continuum model dynamics.
In view of this question, a number of characteristic similarities and
differences between energy- and momentum-conserving discrete
models were highlighted. The momentum conserving Klein-Gordon
system with non-potential background forces discussed here differs
from other path-dependent systems, e.g., having friction and/or AC
drive, in the sense that the viscosity and external forces are not
explicitly introduced. This makes the dynamics of the system
peculiar, for example, as it was demonstrated, the existence, the
intensity, and the sign of energy exchange with the surroundings
depends on the symmetry and other characteristics of the motion.
Further investigation of the intriguing dynamic properties of such
non-potential models is important, given the relevance of
path-dependent forces in various applications such as, e.g.,
aerodynamic and hydrodynamic forces, the forces induced in
automatic control systems and others. Such studies are in progress
and will be reported in future publications.



\end{document}